\documentclass[twocolumn,aps,showkeys,amsmath,amssymb]{revtex4}
\usepackage{graphicx}
\usepackage{dcolumn}
\usepackage{amsmath}%
\usepackage{amssymb}%

\begin{document}


\title{Derivation of Coulomb's law based on a mechanical model of electromagnetic field and a spherical source and sink model of electric charges}

\author{Xiao-Song Wang}
\affiliation{Institute of Mechanical and Power Engineering, Henan Polytechnic University,
Jiaozuo, Henan Province, 454000, China}
\date{Jan. 22, 2021}

\begin{abstract}
We suppose that vacuum is filled with a kind of continuously distributed matter which may be called the $\Omega(1)$ substratum, or the electromagnetic aether. Suppose that the time scale of a macroscopic observer is very
large compares to the the Maxwelllian relaxation time of the $\Omega(1)$ substratum. Thus, the macroscopic observer concludes
 that the $\Omega(1)$ substratum behaves like a Newtonian-fluid. Inspired by H. A. Lorentz, we speculate that electric charges may be extremely small hard spherical sources or spherical sinks with finite radii. Based on the spherical source and spherical sink model of electric charges, we derive Coulomb's law of interactions between static electric charges in vacuum. Further, we derive a reduced form of the Lorentz's force law for static electric charges in vacuum.
\end{abstract}


\maketitle


\section{Introduction \label{sec 100}}
\thispagestyle{plain}

\newtheorem{assumption}{\bfseries Assumption}

\newtheorem{definition}[assumption]{\bfseries Definition}

\newtheorem{lemma}[assumption]{\bfseries Lemma}

\newtheorem{proposition}[assumption]{\bfseries Proposition}

\newtheorem{theorem}[assumption]{\bfseries Theorem}

\newtheorem{wcorollary}[assumption]{\bfseries Corollary}

  Coulomb's law of interactions between static
electric charges in vacuum can be written as \cite{Jackson1963}
\begin{equation}\label{Coulomb's law}
\mathbf{F}_{12}=\frac{1}{4 \pi
\epsilon_{0}}\frac{q_1q_2}{r^2}
\hat{\mathbf{r}}_{12},
\end{equation}
where $q_1$ and $q_2$ are the electric quantities of two electric
charges, $r$ is the distance between the two electric charges,
$\epsilon_{0}$ is the dielectric constant of vacuum,
$\mathbf{F}_{12}$ is the force exerted on the
electric charge with electric quantity $q_2$
 by the electric
charge with electric quantity $q_1$,
 $\hat{\mathbf{r}}_{12}$
denotes the unit vector directed outward along the line from the
electric charge with electric quantity $q_1$ to the electric charge
with electric quantity $q_2$.

  The main purpose of this manuscript is to derive Coulomb's law of interactions between static
electric charges in vacuum by means of fluid mechanics based on
spherical source and spherical sink model of particles.

  The motive of this manuscript is to seek a mechanism of Coulomb's law.
  The reasons why new mechanical interpretations
 of Coulomb's law are interesting may be summarized as follows.

Firstly, Coulomb's law is an elementary law in physics and play various roles in the fields of electromagnetism, electrodynamics, quantum mechanics, cosmology and thermodynamics, etc. \cite{Spavieri2004,Tu-Jun2004}.
From the point view of reductionism, the fundamental importance of Coulomb's law in all branches of
physics urges the reductionists to provide it a proper mechanical interpretation.

Secondly, the mechanism of this action-at-a-distance Coulomb's law remains an
unsolved problem in physics for more than 200 years after the law
was put forth by Coulomb in 1785
\cite{Cavendish1879,WhittakerE1953,Tu-Jun2004,Spavieri2004,Sidharth2005}.
A satisfactory mechanical interpretation in the framework of Descartes' scientific
research program \cite{WhittakerE1953} is interesting.

Thirdly, although the Maxwell's theory of electromagnetic phenomena
is a field theory \cite{Jackson1963}, the concept of field is
different from that of continuum mechanics
\cite{Truesdell1966,Landau-Lifshitz1975,Landau-Lifshitz1975,Fung1977,Eringen1980}
because of the absence of a continuum. Thus, the Maxwell's theory
can only be regarded as a phenomenological theory. New mechanical
interpretations
 of Coulomb's law may help us to establish a field theory of electromagnetic phenomena \cite{Green1962,Jackson1963}.

    Fourthly, there exist some inconsistencies and inner difficulties in the classical
   electrodynamics \cite{Landau-Lifshitz1975,Dirac1978,Whitney1988,Chubykalo-Smirnov-Rueda1996}.
   New theories of Coulomb's law may help to resolve such difficulties.

   Finally, one of the tasks of physics is the unification of the four
fundamental interactions in the universe. New theories of
interactions between static electric charges may shed some light on
this puzzle.

   To conclude, it seems that new considerations on Coulomb's law is needed.

  In this manuscript, we show that Coulomb's law of interactions between static electric charges may be derived based on a mechanical model of vacuum and a spherical source and sink model of electric charges.

\section{A brief introduction of a mechanical model of electromagnetic field \label{sec 200}}
Maxwell's equations in vacuum can be written as \cite{Jackson1963}
\begin{eqnarray}
\nabla \cdot \mathbf{E}&\!\!=\!\!&\frac{\rho_{e}}{\epsilon_{0}}\,, \label{Maxwell 150-110} \\
\nabla \times\mathbf{E} &\!\!=\!\!& -\frac{\partial \mathbf{B}}{\partial t}\,, \label{Maxwell 150-120} \\
\nabla \cdot \mathbf{B}&\!\!=\!\!& 0\,, \label{Maxwell 150-130} \\
\frac{1}{\mu_{0}}\,\nabla \times\mathbf{B}
 &\!\!=\!\!& \mathbf{j}_{e}+\epsilon_{0}\,\frac{\partial \mathbf{E}}{\partial t} \,, \label{Maxwell 150-140}
\end{eqnarray}
where $\mathbf{E}$ is the electric field vector,
$\mathbf{B}$ is the magnetic induction vector,
$\rho_{e}$ is the density field of electric charges,
$\mathbf{j}_{e}$ is the electric current density, $\epsilon_{0}$
is the dielectric constant of vacuum, $\mu_{0}$ is magnetic
permeability of vacuum, $t$ is time, $\nabla =
\mathbf{i}\partial /\partial x +
\mathbf{j}\partial /\partial y +
\mathbf{k}\partial /\partial z$ is the Hamilton operator.

In 1846, W.\ Thomson compared electric phenomena with elasticity. He pointed out that the elastic displacement
$\mathbf{u}$ of an incompressible elastic solid is a possible analogy to the vector electromagnetic potential
$\mathbf{A}$  \cite{WhittakerE1953}. In 1845--1862, G.\ G.\ Stokes suggested that the electromagnetic aether might behave like a glue-water jelly  \cite{Stokes1845,Stokes1849,Stokes1862}.

Following G. G. Stokes, we proposed a  visco-elastic continuum model of vacuum in 2008 \cite{WangXS200804}. Maxwell's equations (\ref{Maxwell 150-110}-\ref{Maxwell 150-140}) in vacuum are derived by methods of continuum mechanics based on a mechanical model of vacuum and a source or sink flow model of electric charges \cite{WangXS200804}. The $\Omega(1)$ substratum behaves as a visco-elastic continuum \cite{WangXS200804}. Maxwell's equations approximate the macroscopic behavior of the $\Omega(1)$ particles, in analogy to the way that classical elastic mechanics approximates the macroscopic behavior of the atoms of solid materials. We briefly review this mechanical model \cite{WangXS200804} of electromagnetic field in this section.

\begin{assumption}\label{p050522 substratum 300-50}
Suppose that vacuum is filled with a kind of continuously distributed matter, which may be called the $\Omega(1)$ substratum. Suppose that all the mechanical quantities of the $\Omega(1)$ substratum under consideration, such as the density, displacements, strains, stresses, etc., are piecewise continuous functions of space and time. Further, we suppose that the material points of the $\Omega(1)$ substratum remain be in one-to-one correspondence with the
material points before a deformation appears.
\end{assumption}

In order to describe the deformation of the $\Omega(1)$ substratum, we introduce a Cartesian coordinate system $\{ 0\rule{-.5pt}{0pt}, \rule{-.5pt}{0pt}x, \rule{-.5pt}{0pt}y\rule{-.5pt}{0pt}, \rule{-.5pt}{0pt}z \rule{-.5pt}{0pt}\}$ or $\{ 0, x_1, x_2, x_3 \}$ which is attached to the $\Omega(1)$ substratum.
\begin{assumption}\label{p050522 homogeneous 300-100}
Suppose that the material of the $\Omega(1)$ substratum under consideration is homogeneous, that is
\begin{equation}\label{p050522 notation 800-200}
\frac{\partial\rho_{1}}{\partial x}=\frac{\partial\rho_{1}}{\partial y}
=\frac{\partial\rho_{1}}{\partial z}=\frac{\partial\rho_{1}}{\partial t}=0,
\end{equation}
where $\rho_{1}$ is the density of the $\Omega(1)$ substratum.
\end{assumption}

\begin{assumption}\label{p050522 isothermal 300-300}
Suppose that the deformation processes of the $\Omega(1)$ substratum are isothermal. So we neglect the thermal effects. Suppose that the deformation processes are not influenced by the gradient of the stress tensor.
Suppose that the material of the $\Omega(1)$ substratum under consideration is isotropic. Supposee that the deformation of the $\Omega(1)$ substratum under consideration is small. Suppose that there are no initial stress and strain in the body under consideration.
\end{assumption}

We introduce the following assumption \cite{WangXS200804}.
\begin{assumption}\label{p050522 constitutive 300-3400}
Suppose the constitutive relation of the $\Omega(1)$ substratum
satisfies the following relationships
\begin{equation}\label{p050522 constitutive 300-2800}
\frac{d e_{ij}}{d t} = \frac{1}{2\eta}\,s_{ij} + \frac{1}{2W} \frac{d s_{ij}}{d t},
\end{equation}
where $e_{ij}$ is the strain deviator, $t$ is time, $s_{ij}$ is the stress deviator, $\eta$ is the dynamic viscocity, $W$ is the shear modulus.
\end{assumption}

We call the materials behaving like Eq.\,(\ref{p050522 constitutive 300-2800}) as Maxwell liquids since J. C. Maxwell
established such a constitutive relation in 1868 \cite{Maxwell1868,Reiner1960,Christensen1982,Joseph1990}.
We introduce the following definition of Maxwellian relaxation time $\tau$ \cite{WangXS200804}
\begin{equation}\label{p050522 relaxation 300-3200}
\tau = \frac{\eta}{W}\,.
\end{equation}

Using Eq.\,(\ref{p050522 relaxation 300-3200}), Eq.\,(\ref{p050522 constitutive 300-2800}) can be written as \cite{WangXS200804}
\begin{equation}\label{p050522 constitutive 300-3300}
 \frac{s_{ij}}{\tau} + \frac{d s_{ij}}{d t} = 2 W\, \frac{d e_{ij}}{d t}\,.
\end{equation}

The vectorial form of the equation of momentum conservation of the $\Omega(1)$ substratum can be written as  \cite{Pao-Mow1973,Eringen1975,Fung1977,Wang1982,Pao1983,Landau-Lifshitz1986}
\begin{equation}\label{p050522 momentum 300-4200}
 W \nabla^2 \mathbf{u}
 + (W +\lambda )\nabla (\nabla \cdot \mathbf{u})
 + \mathbf{f} = \rho_{1}\, \frac{\partial^2 \mathbf{u}}{\partial t^2},
\end{equation}
where $\mathbf{u}$ is the displacement, $\lambda$ is Lam\'{e} constant, $\mathbf{f}$ is the
volume force density exerted on the $\Omega(1)$ substratum, $\nabla^2 = \partial^2
/\partial x^2 + \partial^2 /\partial y^2 +\partial^2 /\partial z^2$ is the Laplace operator.

Let $T_0$ be the characteristic time of a macroscopic observer of an electric charge. We may suppose that the observer's time scale $T_0$ is very large comparing to the the Maxwelllian relaxation time $\tau$. So the Maxwelllian relaxation time $\tau$ is a relatively small number and the stress deviator $s_{ij}$ changes very slowly. Thus, the second term in the left side of Eq.(\ref{p050522 constitutive 300-3300}) may be neglected. According to this macroscopic observer,  the constitutive relation of the $\Omega(1)$ substratum may be written as
\begin{equation}\label{the constitutive relation of aether 4-3}
 s_{ij} = 2 \eta \frac{d e_{ij}}{d t}.
\end{equation}

Therefore, the observer concludes  that the $\Omega(1)$ substratum behaves like the Newtonian-fluid.
We introduce the following definition of point source and sink \cite{WangXS200810}.
Suppose that there exist a singularity at a point $P_0=(x_0,y_0,z_0)$ in a
continuum. If the velocity field of the singularity at a point $P=(x,y,z)$ is
\begin{equation}\label{velocity 300-200}
\mathbf{v}(x,y,z,t)=\frac{Q}{4\pi
r^2}\,\hat{\mathbf{r}},
\end{equation}
where $r=\sqrt{(x-x_0)^2+(y-y_0)^2+(z-z_0)^2}$,
$\hat{\mathbf{r}}$
 is the unit vector directed outward along the line
from the singularity to this point $P=(x,y,z)$, we call such a
singularity a point source in the case of $Q>0$ or a point sink in the case of $Q<0$. Here $Q$ is
called the strength of the source or sink.

For the case of continuously distributed point sources or sinks, it is useful to introduce a definition for the volume density $\rho_s$ of point sources or sinks. The definition is
\begin{equation}\label{definition 300-100}
\rho_s=\lim_{\triangle V \rightarrow 0}\frac{\triangle Q}{\triangle
V}\,,
\end{equation}
where $\triangle V$ is a small volume, $\triangle Q$ is the sum of the strengthes of all the point sources or sinks in the volume $\triangle V$.

The idea that all microscopic particles are sink flows in a fluidic substratum has been proposed by many researchers in the history, for instance, J. C. Maxwell (\cite{WhittakerE1951}, p.\ 243), B. Riemann (\cite{RiemannB2004}, p.\ 507), H. Poincar$\acute{e}$ (\cite{PoincareH1997}, p.\ 171), J. C. Taylor (\cite{Taylor2001}, p.\ 431-436). Therefore, we
suppose that all the electric charges in the universe are the sources or
sinks in the $\Omega(1)$ substratum \cite{WangXS200804}. We define such a source as a
negative electric charge. We define such a sink as a positive electric
charge. The electric charge quantity $q_{e}$ of an electric charge is defined as \cite{WangXS200804}
\begin{equation}\label{charge 300-900}
q_{e} = -\, k_{Q}\rho_{1}\, Q\,,
\end{equation}
where $k_{Q}$ is a positive dimensionless constant.

For the case of continuously distributed electric charges, it is useful to introduce the following definition of the volume density $\rho_e$ of electric charges \cite{WangXS200804}
\begin{equation}\label{charge 300-1000}
\rho_e = \lim_{\triangle V \rightarrow 0}\frac{\triangle
q_{e}}{\triangle V}\,,\rule[-11pt]{0pt}{0pt}
\end{equation}
where $\triangle V$ is a small volume, $\triangle q_{e}$ is the sum
of the strengthes of all the electric charges in the volume $\triangle
V$.

Using Eq.\,(\ref{definition 300-100}-\ref{charge 300-1000}), we have \cite{WangXS200804}
\begin{equation}\label{relation 300-1100}
\rho_e = -\, k_{Q} \rho_{1} \rho_{s}.
\end{equation}

According to Eq.\ (\ref{charge 300-900}) and Eq.\ (\ref{velocity 300-200}), the masses bearing positive electric charges are changing since the strength of a sink evaluates the volume of the $\Omega(1)$ substratum entering the sink per unit of time. Therefore, the equation of mass conservation of the $\Omega(1)$ substratum can be written as \cite{WangXS200804}
\begin{equation}\label{conservation 300-1200}
 \nabla \cdot  \mathbf{v} = -\frac{\rho_{e}}{k_{Q}\rho_{1}},
\end{equation}
where $\mathbf{v}$ is the velocity field of the $\Omega(1)$ substratum, which is defined by $\mathbf{v} = \partial \mathbf{u}/\partial t$.

The momentum of a volume element $\triangle V$ of the $\Omega(1)$ substratum containing continuously distributed electric charges, and moving with an average speed $\mathbf{v}_{e}$, changes. Therefore, the equation of momentum
 conservation  Eq.\,(\ref{p050522 momentum 300-4200})  of the $\Omega(1)$ substratum should be written as \cite{WangXS200804}
\begin{equation}\label{momentum 150-1300}
 W \nabla^2 \mathbf{u}
 + (W +\lambda )\nabla (\nabla \cdot \mathbf{u})
 + \mathbf{f}
 = \rho_{1} \,\frac{\partial^2 \mathbf{u}}{\partial t^2}
 - \frac{\rho_{e}\mathbf{v}_{e}}{k_{Q}}\,.
\end{equation}

In order to simplify Eq.\,(\ref{momentum 150-1300}), we may introduce the following assumption \cite{WangXS200804}.
\begin{assumption}\label{incompressible 150-1400}
Suppose that the $\Omega(1)$ substratum is almost incompressible, i.e.,  we suppose that the volume change coefficient $\theta$ is a sufficient small quantity and varies very slowly in the space so that it can be treated as $\theta = 0$.
\end{assumption}

Based on Assumption \ref{incompressible 150-1400}, we have $\nabla \cdot \mathbf{u}=\theta = 0$.  Therefore, the vectorial form of the equation of momentum  conservation Eq.\,(\ref{momentum 150-1300}) reduces to the following form \cite{WangXS200804}
\begin{equation}\label{momentum 150-1500}
 W \nabla^2 \mathbf{u}+\mathbf{f}
 = \rho_{1}\, \frac{\partial^2 \mathbf{u}}{\partial t^2} - \frac{\rho_{e}\mathbf{v}_{e}}{k_{Q}}\,.
\end{equation}

According to the Stokes-Helmholtz resolution theorem \cite{Pao-Mow1973,Eringen1975}, there exist a scalar function $\psi$ and a vector function  $\mathbf{R}$ such that $\mathbf{u}$ is represented by
\begin{equation}\label{Stokes 150-1600}
 \mathbf{u} = \nabla \psi + \nabla \times
 \mathbf{R}.
\end{equation}

We introduce the definitions \cite{WangXS200804}
\begin{equation}\label{potential 150-1700}
 \nabla\phi = k_{E}\,\frac{\partial}{\partial t}(\nabla \psi)\,,
 \quad \mathbf{A} = k_{E}\,\nabla \times
 \mathbf{R}\,,
\end{equation}
\begin{equation}\label{definition 150-1800}
 \mathbf{E} =
 -\,k_{E}\,\frac{\partial \mathbf{u}}{\partial t}\,, \quad
 \mathbf{B} = k_{E}\, \nabla \times
 \mathbf{u}\,,
\end{equation}
where $\phi$ is the scalar electromagnetic potential, $\mathbf{A}$ is the vector electromagnetic potential, $\mathbf{E}$ is the electric field intensity, $\mathbf{B}$ is the magnetic induction, $k_{E}$ is a positive dimensionless constant.

From Eq.\,(\ref{Stokes 150-1600}), Eq.\,(\ref{potential 150-1700}) and Eq.\,(\ref{definition 150-1800}), we have \cite{WangXS200804}
\begin{equation}\label{field 150-1900}
\mathbf{E} =
 -\nabla \phi - \frac{\partial \mathbf{A}}{\partial
 t}, \quad
 \mathbf{B} = \nabla \times \mathbf{A},
\end{equation}
\begin{eqnarray}
 \nabla \times\mathbf{E} &\!\!\!=\!\!\!&
 -\frac{\partial \mathbf{B}}{\partial t}\,, \label{field 150-2010}\\
 \nabla \cdot \mathbf{B}&\!\!\!=\!\!\!& 0\,.\label{field 150-2020}
 \end{eqnarray}

Noticing $\nabla \cdot \mathbf{u}=0$, $\nabla \cdot \mathbf{A}=0$ and $\mathbf{f}=0$, Eq.\,(\ref{momentum 150-1500}) can be written as \cite{WangXS200804}
\begin{equation}\label{momentum 150-2100}
 \frac{k_{Q}W}{k_{E}} \,\,\nabla \times \mathbf{B} = \frac{k_{Q}\rho_{1}}{k_{E}} \frac{\partial \mathbf{E}}{\partial
 t} +  \rho_{e} \mathbf{v}_{e}\,.
\end{equation}

We introduce the following definitions \cite{WangXS200804}
\begin{equation}\label{p050522 3 defitions}
 \mathbf{j}_{e} = \rho_{e}
 \mathbf{v}_{e}\,, \qquad \epsilon_{0} = \frac{k_{Q}\rho}{k_{E}}\,, \qquad
  \frac{1}{\mu_{0}} = \frac{k_{Q}G}{k_{E}}\,.
\end{equation}

Using Eqs.\,(\ref{p050522 3 defitions}), Eq.\,(\ref{momentum 150-2100}) becomes \cite{WangXS200804}
\begin{equation}\label{momentum 150-2200}
 \frac{1}{\mu_{0}}\,\,\nabla \times\mathbf{B}
 =\mathbf{j}_{e}+\epsilon_{0}\,\frac{\partial \mathbf{E}}{\partial t}\,.
\end{equation}

Noticing Eq.\,(\ref{definition 150-1800}) and Eq.\,(\ref{p050522 3 defitions}), Eq.\,(\ref{conservation 300-1200}) becomes \cite{WangXS200804}
\begin{equation}\label{mass 150-2300}
\nabla \cdot \mathbf{E} = \frac{\rho_{e}}{\epsilon_{0}}\,.
\end{equation}

Eq.\,(\ref{field 150-2010}), Eq.\,(\ref{field 150-2020}), Eq.\,(\ref{momentum 150-2200}) and
Eq.\,(\ref{mass 150-2300}) coincide with Maxwell's equations (\ref{Maxwell 150-110}--\ref{Maxwell 150-140}). Thus, Maxwell's equations (\ref{Maxwell 150-110}--\ref{Maxwell 150-140}) are derived based on this mechanical model of vacuum and the singularity model of electric charges \cite{WangXS200804}.

\section{A Spherical Source and Spherical Sink Model of Electric Charges \label{sec 300}}
Eq.(\ref{the constitutive relation of aether 4-3}) is
the constitutive relation of a Newtonian-fluid.
It is known that the motion of an incompressible Newtonian-fluid is
governed by the Navier--Stokes equations (refer to, for instance,
\cite{Kochin1964,Yih1969,Wu1982a,Landau-Lifshitz1987,Faber1995,Currie2003}),
\begin{equation}\label{Navier--Stokes equations 4-1}
\rho_{1} \left[ \frac{\partial \mathbf{v}}{\partial t}
+ (\mathbf{v} \cdot \nabla )
\mathbf{v} \right ]
 = - \nabla p - \eta \nabla^2 \mathbf{v},
\end{equation}
where $\mathbf{v}$ is the velocity field of the
fluid, $p$ is the pressure field, $\rho_{1}$ is the density field,
$\eta$ is the dynamic viscosity coefficient, $t$ is time.

The definition of Reynolds number $Re$ of a fluid field is
\begin{equation}\label{Reynolds number 4-1}
Re = \frac{\rho_{0} U_{0} L_{0}}{\eta_{0}},
\end{equation}
where $\rho_{0}$ is the characteristic density, $U_0$ is the
characteristic velocity, $L_0$ is the characteristic length,
$\eta_0$ is the characteristic dynamic viscosity coefficient.

\begin{assumption}\label{high Reynolds number}
We speculate that the characteristic velocity $U_0$ of an electric
charge is so high compares to the characteristic dynamic viscosity
coefficient $\eta_0$ of the $\Omega(1)$ substratum that the Reynolds number $Re$ of
the fluid field is a large number.
\end{assumption}

Under this assumption, we may treat the $\Omega(1)$ substratum as an inviscid
incompressible fluid when we study the motion of electric charges.
Therefore, according to Assumption \ref{high Reynolds number}, the
motion of the $\Omega(1)$ substratum is governed by the Euler equations \cite{Kochin1964,Yih1969,Wu1982a,Landau-Lifshitz1987,Faber1995,Currie2003}
\begin{equation}\label{Euler equations 4-1} \frac{\partial
\mathbf{v}}{\partial t} +
(\mathbf{v} \cdot \nabla )
\mathbf{v}
 = - \frac{1}{\rho_{1}}\nabla p.
\end{equation}

If there exist a velocity field which is continuous and finite at
all points of the space, with the exception of individual isolated
points, then these isolated points are called velocity singularities
usually. Point spherical sources and spherical sinks are examples of
singularities.

In 1892 \cite{Lorentz1936}, Lorentz established an electromagnetic theory in order to derive the Fresnel convection coefficient. There are only two types of entities in Lorentz's theory: movable electrons and a stagnant aether. To avoid singularities, the electrons were not designed to be singularities as Larmor's electrons in the aether field, but were extremely small hard spheres with a finite radius.

Inspired by Lorentz \cite{Lorentz1936}, we speculate that electric charges may not be singularities, but may be extremely small hard spherical sources or spherical sinks with finite radii. Thus, we introduce a mechanical model of spherical sources and spherical sinks with finite radii in fluids.

\begin{definition}\label{spherical sources and spherical sinks}
Suppose that there exist a hard sphere with a finite radius $a$ at point
$P_0=(x_0,y_0,z_0)$.
 If the velocity field near the hard sphere at point $P=(x,y,z)$ is
\begin{equation}\label{velocity field of spherical source or spherical sink}
\mathbf{v}(x,y,z,t)=\frac{Q}{4\pi
r^2}\hat{\mbox{\upshape\bfseries{r}}},
\end{equation}
where $r=\sqrt{(x-x_0)^2+(y-y_0)^2+(z-z_0)^2}, r \geq a, $ is the
distance between the point $P_0$ and the point $P$,
$\hat{\mbox{\upshape\bfseries{r}}}$
 denotes the unit vector directed outward along the line
from the point $P_0$ to the point $P$, then we call this hard sphere
a spherical source if $Q>0$ or a spherical sink if $Q<0$. $Q$ is
called the strength of the spherical source or the spherical sink.
\end{definition}

For convenience, we may regard a spherical sink as a negative
spherical source.
Suppose that a static spherical source with strength $Q$ and a radius $a$
locates at the origin $(0, 0, 0)$. In order to calculate the volume
leaving the source per unit time, we enclose the source with an
arbitrary spherical surface $S$ with a radius $b > a$. A calculation
shows that
\begin{equation}\label{volume leaving the source 2-10}
\int\hspace{-1.95ex}\int_{S} \hspace{-3.35ex}\bigcirc \
\mathbf{v}\cdot\mathbf{n}dS
 = \int\hspace{-1.95ex}\int_{S}
\hspace{-3.35ex}\bigcirc \ \frac{Q}{4\pi
b^2}\hat{\mbox{\upshape\bfseries{r}}}\cdot\mathbf{n}dS
= Q,
\end{equation}
where $\mathbf{n}$ denotes the unit vector directed
outward along the line from the origin of the coordinates to the
field point$(x,y,z)$.

Eq. (\ref{volume leaving the source 2-10}) shows that the strength $Q$ of a spherical source or
spherical sink evaluates the volume of the fluid leaving or entering
a control surface per unit time.

Based on the definitions of spherical sources and spherical sinks, we introduce the following model of electric charges.
\begin{assumption}\label{spherical source and spherical sink}
Suppose that all the electric charges in the universe are small hard
spherical sources or spherical sinks with finite radii in the $\Omega(1)$ substratum. We define a spherical source as a negative electric charge.
We define a spherical sink as a positive electric charge. The
electric charge quantity of an electric charge is defined as
\begin{equation}\label{definition of electric charge quantity 4-10}
q_{e} = - k_{Q}\rho_{1} Q,
\end{equation}
where $\rho_{1}$ is the density of the $\Omega(1)$ substratum, $k_{Q}$ is a positive
dimensionless constant, $Q$ is called the strength of the spherical
source or spherical sink.
\end{assumption}

A calculation shows that the mass $m$ of a electric charge is
changing with time as
\begin{equation}\label{the mass is changing}
\frac{dm}{dt} = -\rho_{1} Q = \frac{q_{e}}{k_{Q}},
\end{equation}
where $q_{e}$ is the electric charge quantity of the electric charge.

\section{Forces acting on spherical sources and spherical sinks in ideal fluids \label{sec 400}}
Suppose the velocity field $\mathbf{v}$ of an ideal
fluid is irrotational, then we have \cite{Lamb1932,Kochin1964,Yih1969,Wu1982a,Landau-Lifshitz1987,Faber1995,Currie2003},
\begin{equation}\label{velocity}
\mathbf{v}=\nabla\phi ,
\end{equation}
where $\phi$ is the velocity potential.

It is known that the equation of mass conservation  of an ideal
fluid becomes Laplace's equation
\cite{Lamb1932,Kochin1964,Yih1969,Wu1982a,Landau-Lifshitz1987,Faber1995,Currie2003},
\begin{equation}\label{Laplace's equation}
\nabla^2\phi=0,
\end{equation}
where $\phi$ is velocity potential.

Using spherical coordinates$(r,\theta,\varphi)$, a general form of
solution of Laplace's equation Eq.\ (\ref{Laplace's equation}) can be
obtained by sepatation of variables as \cite{Currie2003}
\begin{equation}\label{solution of Laplace's equation}
\phi(r,\theta)=\sum^{\infty}_{l=0}\left(
A_lr^l+\frac{B_l}{r^{l+1}}\right)P_l(\cos \theta),
\end{equation}
where $A_l$ and $B_l$ are arbitrary constants,
 $P_l(x)$ are Legendre's function of the first kind which is defined as
\begin{equation}\label{Legendre's function}
P_l(x)=\frac{1}{2^ll!}\frac{\mathrm{d}^l}{\mathrm{d}x^l}(x^2-1)^l.
\end{equation}

From Eq.\ (\ref{velocity field of spherical source or spherical sink})
and Eq.\ (\ref{solution of Laplace's equation}), we see that the
velocity potential $\phi(r,\theta)$ of a spherical source or
spherical sink is a solution of Laplace's equation (\ref{Laplace's
equation}).

\begin{proposition}\label{proposition 400-1000}
Suppose that (1) the velocity field $\mathbf{v}$ of a
fluid is irrotational, i.e., we have
\begin{math}
\mathbf{v}=\nabla\phi,
\end{math}
where $\phi$ is the velocity potential; (2) there is an arbitrary
closed surface $S$ fixed in the space without any bodies or
singularities inside $S$; (3) the velocity field
$\mathbf{v}$ is continuous in the closed surface
$S$. Then, we have
\begin{equation}\label{momentum 3-2}
\frac{\mathrm{D}}{\mathrm{D}t} \int\hspace{-1.95ex}\int_{S}
\hspace{-3.35ex}\bigcirc \ \rho_{1} \phi\mathbf{n}dS
=\frac{\partial}{\partial t} \int\hspace{-1.95ex}\int_{S}
\hspace{-3.35ex}\bigcirc \ \rho_{1}\phi\mathbf{n}dS+
\int\hspace{-1.95ex}\int_{S} \hspace{-3.35ex}\bigcirc \
\rho_{1}\mathbf{v}(\mathbf{v}\cdot\mathbf{n})dS.
\end{equation}
where $\mathrm{D} / \mathrm{D}t$ represents the material derivative
in the lagrangian system.
\end{proposition}

For the proof of Proposition \ref{proposition 400-1000}, refer to, for instance,
Appendix 2 in \cite{Wu1982b}.

\begin{theorem}\label{force exerted on spherical sources or spherical sinks by fluids}
Suppose that (1) there exist an ideal fluid  (2) the ideal fluid is
irrotational and barotropic, (3) the density $\rho_{1}$ is homogeneous,
that is
\begin{math}
\partial\rho_{1}/\partial x=\partial\rho_{1}/\partial y
=\partial\rho_{1}/\partial z=\partial\rho_{1}/\partial t=0,
\end{math}
(4) there are no external body forces exerted on the fluid, (5) the
fluid is unbounded and the velocity of the fluid at the infinity is
approaching to zero. Suppose a spherical source or spherical sink is
stationary and is immersed in the ideal fluid. Then, there is a
force
\begin{equation}\label{force on the spherical source 3-1}
\mathbf{F}_Q= \rho_{1}  Q\mathbf{v}_0 + \frac{4 \pi \rho_{1} a^{3}}{3}\frac{\partial
\mathbf{v}_0}{\partial t}
\end{equation}
exerted on the spherical source or the spherical sink by the fluid,
where $\rho_{1}$ is the density of the fluid, $Q$ is the strength of the
spherical source or the spherical sink, $a$ is the radius of the
spherical source or the spherical sink,
$\mathbf{v}_0$ is the velocity of the fluid at the
location of the spherical source induced by all means other than the
spherical source itself.
\end{theorem}

\mbox{\upshape\bfseries{Proof.}}
   Only the proof of the case of a spherical source is needed.
Let us select the coordinates $\{ x, y, z \}$ or $\{ x_1, x_2,
x_3 \}$ that is attached to the static fluid at the infinity.

   We set the origin of the coordinates at the center of the spherical source.
Let $S_{1}$ denotes the spherical surface of the spherical source.
We surround the spherical source by an arbitrary spherical surface
$S_{2}$ with radius $R$ centered at the center of the spherical
source. The outward unit normal to the spherical surface $S_{1}$ and
$S_{2}$ is denoted by $\mbox{\bfseries{n}}$.
   Let $\tau(t)$ denotes the mass system of fluid enclosed in the volume
between the surface $S_{1}$ and the surface $S_{2}$ at time $t$.

   Let $\mathbf{F}_Q$ denotes the hydrodynamic force
exerted on the spherical source by the mass system $\tau$. Then
according to Newton's third law, a reacting force of the force
$\mathbf{F}_Q$ must act on the the fluid enclosed
in the mass system $\tau$. Let $\mathbf{F}_1 = -
\mathbf{F}_Q$ denotes this reacting force acted on
the mass system $\tau$ by the spherical source through the surface
$S_{1}$. Let $\mathbf{F}_2$ denotes the
hydrodynamic force exerted on the mass system $\tau$ due to the
pressure distribution on the surface $S_{2}$.
 Let $\mbox{\bfseries{K}}$ denotes momentum of the mass system $\tau$.

Applying Newton's second law of motion to the mass system $\tau$, we have
\begin{equation}\label{Newton's second law of motion}
\frac{\mathrm{D}\mathbf{K}}{\mathrm{D}t}
=\mathbf{F}_1 + \mathbf{F}_2
=-\mathbf{F}_Q + \mathbf{F}_2,
\end{equation}
where $\mathrm{D} / \mathrm{D}t$ represents the material derivative
in the lagrangian system (see, for instance,
\cite{Kochin1964,Yih1969,Wu1982a,Landau-Lifshitz1987,Faber1995,Currie2003}).

In order to calculate $\mathbf{F}_Q$, we
calculate $\mathrm{D}\mathbf{K}/\mathrm{D} t$ and
$\mathbf{F}_2$ respectively. The expressions of the
momentum $\mathbf{k}$ is
\begin{equation}\label{integral 300-1000}
\mathbf{K}
  =  \int\hspace{-1.5ex}\int\hspace{-1.5ex}\int_\tau \rho_{1}\mathbf{v}dV,
\end{equation}
where the integral is a volume integral, $\mathbf{n}$ denotes the unit vector directed outward
along the line from the origin of the coordinates to the field
point$(x,y,z)$.

The expressions of the force $\mathbf{F}_2$ is
\begin{equation}\label{integral 300-1100}
\mathbf{F}_2  = \int\hspace{-1.95ex}\int_{S_2} \hspace{-4.15ex}\bigcirc -p \mathbf{n}dS,
\end{equation}
where the integral is a surface integral.

Since the velocity field is irrotational, we have the following
relation
\begin{equation}\label{irrotational velocity field 3-20}
\mathbf{v}=\nabla\phi,
\end{equation}
where $\phi$ is the velocity potential.

According to Ostrogradsky--Gauss theorem (see, for instance,
\cite{Kochin1964,Yih1969,Wu1982a,Faber1995,Currie2003}) and using
Eq.\ (\ref{irrotational velocity field 3-20}), we have
\begin{eqnarray}\label{Ostrogradsky-Gauss 3-3}
\int\hspace{-1.5ex}\int\hspace{-1.5ex}\int_{\tau}
\rho_{1}\mathbf{v}dV
&=&\int\hspace{-1.5ex}\int\hspace{-1.5ex}\int_{\tau}
\rho_{1}\nabla\phi dV \nonumber \\
&=&\int\hspace{-1.95ex}\int_{S_2} \hspace{-4.15ex}\bigcirc \
\rho_{1}\phi\mathbf{n}dS -
\int\hspace{-1.95ex}\int_{S_1} \hspace{-4.15ex}\bigcirc \
\rho_{1}\phi\mathbf{n}dS.
\end{eqnarray}

Using Eq.\ (\ref{integral 300-1000}) and Eq.\ (\ref{Ostrogradsky-Gauss 3-3}), we have
\begin{equation}\label{momentum 3-1}
\frac{\mathrm{D}\mathbf{K}}{\mathrm{D}t}
=\frac{\mathrm{D}}{\mathrm{D}t} \left [
\int\hspace{-1.95ex}\int_{S_2} \hspace{-4.15ex}\bigcirc \
\rho_{1}\phi\mathbf{n}dS -
\int\hspace{-1.95ex}\int_{S_1} \hspace{-4.15ex}\bigcirc \
\rho_{1}\phi\mathbf{n}dS \right ].
\end{equation}

Applying Proposition \ref{proposition 400-1000} to the first integral in Eq.\ (\ref{momentum 3-1}), we have
\begin{equation}\label{momentum 3-250}
\frac{\mathrm{D}}{\mathrm{D}t} \int\hspace{-1.95ex}\int_{S_2}
\hspace{-4.15ex}\bigcirc \ \rho_{1}\phi\mathbf{n}dS
=\frac{\partial}{\partial t} \int\hspace{-1.95ex}\int_{S_2}
\hspace{-4.15ex}\bigcirc \ \rho_{1}\phi\mathbf{n}dS +
\int\hspace{-1.95ex}\int_{S_2} \hspace{-4.15ex}\bigcirc \
\rho_{1}\mathbf{v}(\mathbf{v}\cdot\mathbf{n})dS.
\end{equation}

Putting Eq.\ (\ref{momentum 3-250}) into Eq.\ (\ref{momentum 3-1}), we
have
\begin{eqnarray}\label{momentum 3-3}
\frac{\mathrm{D}\mathbf{K}}{\mathrm{D}t} &=&
\frac{\partial}{\partial t} \int\hspace{-1.95ex}\int_{S_2}
\hspace{-4.15ex}\bigcirc \ \rho_{1}\phi\mathbf{n}dS +
\int\hspace{-1.95ex}\int_{S_2}
\hspace{-4.15ex}\bigcirc \ \rho_{1}\mathbf{v}(\mathbf{v}\cdot\mathbf{n})dS \nonumber \\
&-& \frac{\mathrm{D}}{\mathrm{D}t} \int\hspace{-1.95ex}\int_{S_1}
\hspace{-4.15ex}\bigcirc \ \rho_{1}\phi\mathbf{n}dS.
\end{eqnarray}

Now, we calculate $\mathbf{F}_2$. According to
Lagrange--Cauchy integral (see, for instance,
\cite{Kochin1964,Yih1969,Wu1982a,Faber1995,Currie2003}), we have
\begin{equation}\label{Lagrange-Cauchy integral}
\frac{\partial\phi}{\partial t}+
\frac{(\nabla\phi)^2}{2}+\frac{p}{\rho_{1}}=f(t),
\end{equation}
where $f(t)$ is an arbitrary function of time $t$. Since the
velocity $\mathbf{v}$ of the fluid at the infinity
is approaching to zero,
 and noticing Eq.\  (\ref{solution of Laplace's equation}),
 $\phi(t)$ must be of the following form
\begin{equation}
\phi(r,\theta,t)=\sum^{\infty}_{l=0} \frac{B_l(t)}{r^{l+1}}P_l(\cos
\theta),
\end{equation}
where $B_l(t), l \geq 0$ are functions of time $t$.

Thus, we have the following estimations at the infinity of the velocity field
\begin{equation}\label{estimation}
\phi =O\left( \frac{1}{r}\right), \quad \frac{\partial\phi}{\partial
t} =O\left( \frac{1}{r}\right), \quad r \rightarrow \infty,
\end{equation}
where $\varphi(x) = O(\psi(x)), x \rightarrow a$ stands for
$\overline{\lim}_{x \rightarrow a} \mid\varphi(x)\mid / \psi(x) = k,
(0 \leq k < +\infty).$

Applying Eq.\  (\ref{Lagrange-Cauchy integral}) at the infinity and using
Eq.\  (\ref{estimation}), we have
$\mid\mathbf{v}\mid\rightarrow0$,
 $\partial\phi / \partial t\rightarrow0$ and $p=p_{\infty}$,
 where $p_{\infty}$ is a constant. Thus,
 $f(t)=p_\infty / \rho_{1}$. Therefore, according to (\ref{Lagrange-Cauchy integral}), we have
\begin{equation}\label{the pressure distribution}
p=p_\infty-\rho_{1}\frac{\partial\phi}{\partial t}-
\frac{\rho_{1}(\mathbf{v}\cdot\mathbf{v})}{2}.
\end{equation}

Using Eq.\  (\ref{integral 300-1100}) and Eq.\  (\ref{the pressure distribution}), we have
\begin{equation}\label{force due to the surface S2 3-1}
\mathbf{F}_2 =\int\hspace{-1.95ex}\int_{S_2}
\hspace{-4.15ex}\bigcirc \ \rho_{1}\frac{\partial\phi}{\partial
t}\mathbf{n}dS
 +\int\hspace{-1.95ex}\int_{S_2} \hspace{-4.15ex}\bigcirc \
\frac{\rho_{1}(\mathbf{v}\cdot\mathbf{v})\mathbf{n}}{2}dS.
\end{equation}

Putting Eq.\  (\ref{momentum 3-3}) and Eq.\  (\ref{force due to the surface
S2 3-1}) into Eq.\  (\ref{Newton's second law of motion}), we have
\begin{eqnarray}\label{force on the source 3-1}
\mathbf{F}_Q &=&
\int\hspace{-1.95ex}\int_{S_2}\hspace{-4.15ex}\bigcirc \
\left[\frac{1}{2}\rho_{1}(\mathbf{v}\cdot\mathbf{v})\mathbf{n}
-\rho_{1}\mathbf{v}(\mathbf{v}\cdot\mathbf{n})\right]dS \nonumber \\
&+& \frac{\mathrm{D}}{\mathrm{D}t} \int\hspace{-1.95ex}\int_{S_1}
\hspace{-4.15ex}\bigcirc \ \rho_{1}\phi\mathbf{n}dS.
\end{eqnarray}

Since the radius $R$ of the spherical surface $S_{2}$ is arbitrary,
we may let $R$ to be large enough. Applying the result (5.13) in \cite{Currie2003}, we have
\begin{equation}\label{force due to the surface S2 3-3}
\int\hspace{-1.95ex}\int_{S_2}
\hspace{-4.15ex}\bigcirc \
\left[\frac{1}{2}\rho_{1}(\mathbf{v}\cdot\mathbf{v})\mathbf{n}
-\rho_{1}\mathbf{v}(\mathbf{v}\cdot\mathbf{n})\right]dS=0.
\end{equation}

Thus, using Eq.\  (\ref{force due to the surface S2 3-3}),
Eq.\  (\ref{force on the source 3-1}) becomes
\begin{equation}\label{force on the source 3-2}
\mathbf{F}_Q = \frac{\mathrm{D}}{\mathrm{D}t}
\int\hspace{-1.95ex}\int_{S_1} \hspace{-4.15ex}\bigcirc \
\rho_{1}\phi\mathbf{n}dS.
\end{equation}

Applying Proposition \ref{proposition 400-1000}, Eq.\  (\ref{force on
the source 3-2}) becomes
\begin{equation}\label{force on the source 3-3}
\mathbf{F}_Q = \frac{\partial}{\partial t}
\int\hspace{-1.95ex}\int_{S_1} \hspace{-4.15ex}\bigcirc \
\rho_{1}\phi\mathbf{n}dS +
\int\hspace{-1.95ex}\int_{S_1} \hspace{-4.15ex}\bigcirc \
\rho_{1}\mathbf{v}(\mathbf{v}\cdot\mathbf{n})dS.
\end{equation}

For convenience, we introduce the following definitions
\begin{equation}
I_1 = \frac{\partial}{\partial t} \int\hspace{-1.95ex}\int_{S_1}
\hspace{-4.15ex}\bigcirc \ \rho_{1}\phi\mathbf{n}dS, \label{integral 400-1000}
\end{equation}
\begin{equation}
I_2 = \int\hspace{-1.95ex}\int_{S_1} \hspace{-4.15ex}\bigcirc \
\rho_{1}\mathbf{v}(\mathbf{v}\cdot\mathbf{n})dS. \label{integral 400-1100}
\end{equation}

Thus, Eq.\ (\ref{force on the source 3-3}) becomes
\begin{equation}\label{force on the source 3-4}
\mathbf{F}_Q = I_1 + I_2.
\end{equation}

Now we calculate the two terms in Eq.\ (\ref{force on the source
3-4}) respectively. Firstly, we calculate the integral $I_1$ in
Eq.\ (\ref{integral 400-1000}). Since the velocity field
induced by the spherical source with strength $Q$ is
Eq.\ (\ref{velocity field of spherical source or spherical sink}),
then according to the superposition principle
 of velocity field of ideal fluids, the velocity $\mathbf{v}$ on the surface $S_1$ is
\begin{equation}\label{velocity on the surface S1 3-1}
\mathbf{v} =\frac{Q}{4\pi
a^2}\mathbf{n} + \mathbf{v}_0,
\end{equation}
where $\mathbf{n}$ denotes the unit vector directed
outward.

Since the velocity field $\mathbf{v}$ is
irrotational, we have
\begin{equation}\label{velocity on the surface S1 3-2}
\phi = -\frac{Q}{4\pi a} + \phi_0,
\end{equation}
where $\phi_0$ is the velocity potential respect to
$\mathbf{v}_{0}$, i.e.,
$\mathbf{v}_{0}=\nabla\phi_{0}$.

Since the density $\rho_{1}$ is homogeneous, we have
\begin{equation}\label{density is homogeneous 3-1}
\frac{\partial\rho_{1}}{\partial t}=0.
\end{equation}

Noticing Eq.\ (\ref{density is homogeneous 3-1}), Eq.\ (\ref{integral 400-1000}) can be written as
\begin{equation}\label{integral i1 3-1}
I_1 = \int\hspace{-1.95ex}\int_{S_1} \hspace{-4.15ex}\bigcirc \ \rho_{1}
\frac{\partial \phi}{\partial t} \mathbf{n}dS.
\end{equation}

Suppose that $\partial Q / \partial t = 0$. Using
Eq.\ (\ref{velocity on the surface S1 3-2}), we have
\begin{equation}\label{velocity potential on the surface S1 3-1}
\frac{\partial \phi}{\partial t} = \frac{\partial \phi_0}{\partial
t} - \frac{1}{4\pi a}\frac{\partial Q}{\partial t} = \frac{\partial
\phi_0}{\partial t}.
\end{equation}

 Using Eq.\ (\ref{velocity potential on the surface S1 3-1}) and Ostrogradsky--Gauss theorem,
 Eq.\ (\ref{integral i1 3-1}) becomes
\begin{eqnarray}\label{integral i1 3-2}
I_1 &=& \frac{\partial}{\partial t} \int\hspace{-1.95ex}\int_{S_1}
\hspace{-4.15ex}\bigcirc \ \rho_{1} \phi_0 \mathbf{n}dS \nonumber \\
&=& \frac{\partial}{\partial t}
\int\hspace{-1.5ex}\int\hspace{-1.5ex}\int_{V_1}
\rho_{1}\nabla\phi_0 dV \nonumber \\
&=& \frac{\partial}{\partial t}
\int\hspace{-1.5ex}\int\hspace{-1.5ex}\int_{V_1} \rho_{1}
\mathbf{v}_0 dV.
\end{eqnarray}

We speculate that the radius $a$ of the spherical source may be so
small that the velocity $\mathbf{v}_0$ at any point
of the spherical surface of the spherical source may be treated as a constant. Thus,
Eq.\ (\ref{integral i1 3-2}) becomes
\begin{eqnarray}\label{integral i1 3-3}
I_1 &=& \frac{\partial (\rho_{1}
\mathbf{v}_0)}{\partial t}
\int\hspace{-1.5ex}\int\hspace{-1.5ex}\int_{V_1} dV \nonumber \\
&=& \frac{\partial (\rho_{1} \mathbf{v}_0)}{\partial
t} \frac{4 \pi a^{3}}{3} \nonumber \\
&=& \frac{4 \pi \rho_{1} a^{3}}{3}\frac{\partial
\mathbf{v}_0}{\partial t}.
\end{eqnarray}

Now we calculate the integral $I_2$  in Eq.\ (\ref{integral 400-1100}).
Noticing Eq.\ (\ref{velocity on the surface S1 3-1}), we have
\begin{eqnarray}\label{integral i2 3-1}
I_2
 &=& \rho_{1}\int\hspace{-1.95ex}\int_{S_1}
\hspace{-4.15ex}\bigcirc \
 \left[\frac{Q^2}{16\pi^2 a^4}\mathbf{n}
 + \frac{Q}{4\pi a^2}\mathbf{v}_0 \right. \nonumber \\
 &&\left. + \frac{Q}{4\pi a^2}(\mathbf{v}_0\cdot\mathbf{n})\mathbf{n}
 + (\mathbf{v}_0\cdot\mathbf{n})\mathbf{v}_0
 \right ]dS.
\end{eqnarray}

For convenience, we introduce the following definitions
\begin{eqnarray}\label{integral definitions 3-2}
&&J_1 = \int\hspace{-1.95ex}\int_{S_1} \hspace{-4.15ex}\bigcirc \
\frac{\rho_{1} Q^2}{16\pi^2 a^4}\mathbf{n}dS,\nonumber \\
&&J_2 = \int\hspace{-1.95ex}\int_{S_1} \hspace{-4.15ex}\bigcirc \
\frac{\rho_{1} Q}{4\pi a^2}\mathbf{v}_0dS,\nonumber \\
&&J_3 = \int\hspace{-1.95ex}\int_{S_1} \hspace{-4.15ex}\bigcirc \
\frac{\rho_{1} Q}{4\pi a^2}(\mathbf{v}_0\cdot\mathbf{n})\mathbf{n}dS,\nonumber \\
&&J_4 = \int\hspace{-1.95ex}\int_{S_1} \hspace{-4.15ex}\bigcirc \
(\mathbf{v}_0\cdot\mathbf{n})\mathbf{v}_0dS.
\end{eqnarray}

Thus, Eq.\ (\ref{integral i2 3-1}) becomes
\begin{equation}\label{integral i2 3-2}
I_2 = J_1 + J_2 + J_3 + J_4.
\end{equation}

We regard the velocity $\mathbf{v}_0$ at any point of the spherical surface $S_1$ as a constant. Thus,
the four integral terms in Eq.\ (\ref{integral i2 3-2}) turns out to be
\begin{equation}\label{integral j2 3-1}
J_1 = 0, \quad
J_2 = \rho_{1} Q\mathbf{v}_0, \quad
J_3 = 0, \quad
J_4 = 0.
\end{equation}

Thus, using Eq.\ (\ref{integral j2 3-1}), we have
\begin{equation}\label{integral i2 3-3}
I_2 = \rho_{1} Q\mathbf{v}_0.
\end{equation}

Putting Eq.\ (\ref{integral i1 3-3}) and Eq.\ (\ref{integral i2 3-3})
into Eq.\ (\ref{force on the source 3-4}), we obtain
Eq.\ (\ref{force on the spherical source 3-1}).     $\Box$

  Theorem \ref{force exerted on spherical sources or spherical sinks by fluids}
  only considers the situation that the spherical sources or spherical sinks are at rest in fluids.
  Now we consider the case that the spherical sources or spherical sinks are moving in fluid.

\begin{theorem}\label{force exerted on moving
spherical sources or spherical sinks} Suppose that the assumptions
(1),(2),(3),(4) and (5) in Theorem \ref{force exerted on spherical
sources or spherical sinks by fluids} are valid and a spherical
source or a spherical sink is moving in the fluid with a velocity
$\mbox{\upshape\bfseries{v}}_s$, then there is a force
\begin{equation}\label{force on moving source 3-1}
\mathbf{F}_Q= \rho_{1} Q(\mathbf{v}_f-\mbox{\upshape\bfseries{v}}_s) + \frac{4 \pi \rho_{1} a^{3}}{3}\frac{\partial
}{\partial
t}(\mathbf{v}_f-\mbox{\upshape\bfseries{v}}_s)
\end{equation}
is exerted on the spherical source or the spherical sink by the
fluid, where $\rho_{1}$ is the density of the fluid, $Q$ is the strength
of the spherical source or the spherical sink, $a$ is the radius of
the spherical source or the spherical sink,
$\mathbf{v}_f$ is the velocity of the fluid at the
location of the source induced by all means other than the spherical
source itself.
\end{theorem}

\mbox{\upshape\upshape\bfseries{Proof.}}
 The velocity of the fluid relative to the spherical source at the
location of the spherical source is
$\mathbf{v}_f-\mbox{\upshape\bfseries{v}}_s$. Let
us select the coordinates that is attached to the spherical source
and set the origin of the coordinates at the center of the spherical
source. Then Eq.\ (\ref{force on moving source 3-1}) can be obtained
following the same procedures in the proof of Theorem \ref{force
exerted on spherical sources or spherical sinks by
fluids}.    $\Box$

Applying Theorem \ref{force exerted on moving spherical sources or
spherical sinks} to the situation
 that a spherical source or spherical sink is exposed to the velocity field of another spherical source or sink, we have the following result.
\begin{wcorollary}\label{singularity-singularity forces}
Suppose that the assumptions (1),(2),(3),(4) and (5) in Theorem
\ref{force exerted on spherical sources or spherical sinks by
fluids} are valid and a spherical source or spherical sink with
strength $Q_2$ is exposed to the velocity field of another static
spherical source or spherical sink with strength $Q_1$, then the
force $\mathbf{F}_{12}$
 exerted on the spherical source or spherical sink with strength $Q_2$
 by the velocity field of the spherical source or spherical sink with strength $Q_1$ is
\begin{eqnarray}
\mathbf{F}_{12} &=& \rho_{1}
 Q_2 \left ( \frac{Q_1}{4\pi
r^2}\hat{\mathbf{r}}_{12}-\mbox{\upshape\bfseries{v}}_2
\right ) \nonumber \\
&+& \frac{ \rho_{1} a_2^{3}}{3r^2}\frac{\partial \left (Q_1\hat{\mathbf{r}}_{12}-\mbox{\upshape\bfseries{v}}_2
\right )}{\partial t}
,\label{force on moving source 3-2}
\end{eqnarray}
where $\hat{\mathbf{r}}_{12}$ denotes the unit
vector directed outward along the line from the spherical source or
spherical sink with strength $Q_1$ to the spherical source or
spherical sink with strength $Q_2$, $a_2$ is the radius of the
spherical source or the spherical sink with strength $Q_2$, $r$ is
the distance between the two bodies, $\mbox{\upshape\bfseries{v}}_2$
is the moving velocity of the spherical source or spherical sink
with strength $Q_2$.
\end{wcorollary}

If the spherical source with strength $Q_2$ is also static in the $\Omega(1)$ substratum,
then Eq.\ (\ref{force on moving source 3-2}) reduces to
\begin{equation}\label{force on moving source 3-3}
\mathbf{F}_{12}= \frac{\rho_{1}
 Q_1 Q_2}{3\pi
r^2}\hat{\mathbf{r}}_{12}
 + \frac{ \rho_{1} a_2^{3}}{3r^2}\frac{\partial Q_1}{\partial t}
\hat{\mathbf{r}}_{12}.
\end{equation}

\section{Derivation of Coulomb's law of interactions between static electric charges in vacuum \label{sec 600}}
Based on Assumption \ref{spherical source and spherical sink} and Assumption \ref{high Reynolds number},
 we can apply Theorem \ref{force exerted on spherical
sources or spherical sinks by fluids} and Theorem \ref{force exerted
on moving spherical sources or spherical sinks} to study the motions
of electric charges.

\begin{theorem}\label{charge-charge forces}
Suppose that a static electric charge with an electric charge quantity
$q_2$ is exposed to the electric field of another static electric
charge with an electric charge quantity $q_1$, then the force
$\mathbf{F}_{12}$
 exerted on the electric charge with electric charge quantity $q_2$
 by the electric field of the electric charge with
electric charge quantity $q_1$ is
\begin{eqnarray}\label{force on electric charge 5-1}
\mathbf{F}_{12} &=& \frac{1}{k_{E}
k_{Q}}\frac{1}{4 \pi \epsilon_{0}}
 \frac{q_1 q_2}{r^2}\hat{\mathbf{r}}_{12} \nonumber \\
&&-  \frac{a_{2}^{3}}{3 k_{Q}r^2}\frac{\partial q_1}{\partial t} \hat{\mathbf{r}}_{12},
\end{eqnarray}
where $\hat{\mathbf{r}}_{12}$ denotes the unit
vector directed outward along the line from the electric charge with
electric charge quantity $q_1$ to the electric charge with electric
charge quantity $q_2$, $a_{2}$ is the radius of the electric charge
with electric charge quantity $q_2$, $r$ is the distance between the
two electric charges, $k_{Q}$ and $k_{E}$ are two positive
dimensionless constants.
\end{theorem}
\mbox{\upshape\upshape\bfseries{Proof.}}
From Assumption \ref{spherical source and spherical sink}, we have
\begin{equation}\label{strengthes and electric charge quantity 5-1}
Q_1 = - \frac{q_1}{k_{Q}\rho_{1}}, \quad  Q_2 = - \frac{q_2}{k_{Q}\rho_{1}},
\end{equation}where
$Q_1$ and $Q_2$ are the strengthes of the electric charges
respectively. From a definition in \cite{WangXS200804}, we
have
\begin{equation}\label{dielectric constant of vacuum 5-1}
\epsilon_{0} = \frac{k_{Q}\rho_{1}}{k_{E}},
\end{equation}
where $\epsilon_{0}$ is the dielectric constant of vacuum. Putting
Eq.\ (\ref{strengthes and electric charge quantity 5-1}) and
Eq.\ (\ref{dielectric constant of vacuum 5-1}) into Eq.\ (\ref{force on
moving source 3-3}), we obtain Eq.\ (\ref{force on electric charge 5-1}).
$\Box$

If we ignore the second term in right side of Eq.\ (\ref{force on electric charge
5-1}), then we have
\begin{equation}\label{force on electric charge 5-1.5}
\mathbf{F}_{12} = \frac{1}{k_{E} k_{Q}}\frac{1}{4
\pi \epsilon_{0}} \frac{q_1 q_2}{r^2}\hat{\mathbf{r}}_{12}.
\end{equation}

Compare Eq.\ (\ref{force on electric charge 5-1.5}) with
Eq.\ (\ref{Coulomb's law}), it is natural for us to introduce the following
assumption.
\begin{assumption}\label{contants relation}
Suppose we have the following relation
\begin{equation}\label{contants relation 5-1}
k_{E} k_{Q} = 1.
\end{equation}
\end{assumption}

Based on Assumption \ref{contants relation}, Eq.\ (\ref{force on electric charge 5-1.5}) has the same form of Coulomb's
law (\ref{Coulomb's law}) of interactions between static electric charges in vacuum.

Theorem \ref{charge-charge forces} only states the force
 exerted on a static electric charge
 by the electric field of another static electric charge. We may
 generalize this result to the case of an static electric charge
exposed to any electric field of the $\Omega(1)$ substratum.

We introduce the following definition \cite{WangXS200804}
\begin{equation}\label{definition of electric field intensity 5-1}
 \mbox{\upshape\bfseries{E}} =
 -k_{E}\frac{\partial \mathbf{u}}{\partial t},
\end{equation}
 where $\mathbf{u}$ is the
displacement of the visco-elastic aether, $\partial
\mathbf{u}/\partial t$ is the velocity field of
the $\Omega(1)$ substratum, $\mbox{\upshape\bfseries{E}}$ is the electric field
intensity, $k_{E}$ is a positive dimensionless constant.

Since the observer of an electric charge concludes that the $\Omega(1)$ substratum
behaves as a Newtonian-fluid under his time scale, we may define the
electric field intensity as the velocity field of the $\Omega(1)$ substratum.
Therefore, we define the electric field intensity in the $\Omega(1)$ substratum as
\begin{equation}\label{electric 600-1000}
 \mbox{\upshape\bfseries{E}} =
 -k_{E}\mbox{\upshape\bfseries{v}}_{\Omega(1)},
\end{equation}
 where
$\mbox{\upshape\bfseries{v}}_{\Omega(1)}$ is the velocity field of the $\Omega(1)$ substratum
at the location of an testing electric charge induced by all means
other than the testing electric charge itself,
$\mbox{\upshape\bfseries{E}}$ is the electric field intensity,
$k_{E}$ is a positive dimensionless constant.

\begin{theorem}\label{force exerted on a static electric charge 5-1}
Suppose that a static electric charge with an electric charge quantity
$q$ is exposed to an electric field $\mbox{\upshape\bfseries{E}}$ of
the $\Omega(1)$ substratum, then the force $\mathbf{F}$
 exerted on the electric charge
 by the electric field $\mbox{\upshape\bfseries{E}}$ is
\begin{equation}\label{force on electric charge 5-2}
\mathbf{F} = q \mbox{\upshape\bfseries{E}} -
\frac{4 \pi a^{3}\epsilon_0}{3 k_{Q}}\frac{\partial
\mbox{\upshape\bfseries{E}}}{\partial t},
\end{equation}
where $q$ is the electric charge quantity of the electric charge,
$a$ is the radius of the electric charge, $\epsilon_{0}$ is the
dielectric constant of vacuum, $k_{Q}$ is a positive dimensionless
constant, $\mbox{\upshape\bfseries{E}}$ is the electric field at the
location of the electric charge induced by all means other than the
electric charge itself.
\end{theorem}
\mbox{\upshape\upshape\bfseries{Proof.}}
From Assumption \ref{spherical source and spherical sink}, we have
\begin{equation}\label{strengthes and electric charge quantity 5-2}
Q = - \frac{q}{k_{Q}\rho_{1}},
\end{equation}
where $Q$ is the strength of the electric charge. Putting
Eq.\ (\ref{electric 600-1000}) and
Eq.\ (\ref{strengthes and electric charge quantity 5-2}) into
Eq.\ (\ref{force on the spherical source 3-1}) and using
Eq.\ (\ref{dielectric constant of vacuum 5-1}) and Eq.\ (\ref{contants
relation 5-1}), we obtain Eq.\ (\ref{force on electric charge 5-2}).
$\Box$

If we ignore the second term in Eq.\ (\ref{force on electric charge
5-2}), then we have
\begin{equation}\label{force on electric charge 5-3}
\mathbf{F} = q \mbox{\upshape\bfseries{E}}.
\end{equation}

The Lorentz's force law can be written as \cite{Jackson1963}
\begin{equation}\label{force on electric charge 5-4}
\mathbf{F} = q \mbox{\upshape\bfseries{E}} + q
\mbox{\upshape\bfseries{v}}_e \times \mbox{\upshape\bfseries{B}},
\end{equation}
where $q$ is the electric charge quantity of a electric charge,
$\mathbf{F}$ is the force exerted on the electric
charge, $\mbox{\upshape\bfseries{v}}_e$ is the velocity of the
electric charge, $\mbox{\upshape\bfseries{E}}$ and
$\mbox{\upshape\bfseries{B}}$ are the electric field intensity and
the magnetic induction respectively.

Eq.\ (\ref{force on electric charge 5-3}) is a reduced
form of the Lorentz's force law (\ref{force on electric charge 5-4})
for static electric charges in vacuum.

\section{Discussion \label{Discussion}}
According to Eq.\ (\ref{electric 600-1000}), the electric field intensity is a linear function of the
velocity field of the $\Omega(1)$ substratum. Therefore, the superposition principle
of electric fields of static electric charges in vacuum is deduced
from the superposition theorem of the velocity field of fluids.

It is an interesting task to generalize this work further to describe the forces exerted on a
electric charge moving with a velocity in a electromagnetic field.

\section{Conclusion \label{Conclusion}}
Following J. C. Maxwell, we suppose that vacuum is filled with a kind of continuously distributed matter which may be called the $\Omega(1)$ substratum, or the electromagnetic aether.
Suppose that the time scale of a macroscopic observer's is very
large compares to the the Maxwelllian relaxation time of the $\Omega(1)$ substratum. Thus, the macroscopic observer concludes that the $\Omega(1)$ substratum behaves like a Newtonian-fluid.
Inspired by H. A. Lorentz, we speculate that electric charges may not be singularities, but may be extremely small hard spherical sources or spherical sinks with finite radii. Based on the spherical source and spherical sink model of
electric charges, we derive Coulomb's law of interactions between
static electric charges in vacuum. Further, we define the
electric field intensity as a linear function of velocity field of
the $\Omega(1)$ substratum at the location of an testing electric charge induced by
all means other than the testing electric charge itself. Then, we
obtain a reduced form of the Lorentz's force law for static electric
charges in vacuum.

\end{document}